\documentclass[sigconf]{acmart}

\usepackage{comment}
\usepackage{amsmath,amsfonts}
\usepackage{algorithmic}
\usepackage{graphicx}
\usepackage{textcomp}
\usepackage{tabu}
\usepackage{tikz}
\usepackage{pgfplots}
\usepackage{multirow}
\usepackage{booktabs}
\usepackage{makecell}
\usepackage{colortbl}
\usepackage{hhline}
\usepackage{hyperref}
\usepackage{booktabs}
\usepackage{rotating}
\usepackage{longtable}
\usepackage{comment}
\usepackage{enumerate}
\usepackage{enumitem}
\usepackage[utf8]{inputenc}

\AtBeginDocument{%
  \providecommand\BibTeX{{%
    \normalfont B\kern-0.5em{\scshape i\kern-0.25em b}\kern-0.8em\TeX}}}

\setcopyright{acmcopyright}
\copyrightyear{2022}
\acmYear{2022}
\acmDOI{XXXXXXX.XXXXXXX}

\acmConference[EASE '22]{International Conference on Evaluation and Assessment in Software Engineering}{June 13--15,
  2022}{Gothenburg, Sweden}

\copyrightyear{2022} 
\acmYear{2022} 
\setcopyright{acmcopyright}\acmConference[EASE 2022]{The International Conference on Evaluation and Assessment in Software Engineering 2022}{June 13--15, 2022}{Gothenburg, Sweden}
\acmBooktitle{The International Conference on Evaluation and Assessment in Software Engineering 2022 (EASE 2022), June 13--15, 2022, Gothenburg, Sweden}
\acmPrice{15.00}

\begin{document}

\title[Conversational DevBots for Secure Software Development]{Conversational DevBots for Secure Programming: An Empirical Study on SKF Chatbot} 


\author{Catherine Tony, Mohana Balasubramanian, Nicol\'{a}s E. D\'{i}az Ferreyra, and Riccardo Scandariato}
\affiliation{%
  \institution{
  Hamburg University of Technology}
  \city{Hamburg}\country{Germany}
  }
\email{{catherine.tony, mohana.balasubramanian, nicolas.diaz-ferreyra, riccardo.scandariato}@tuhh.de}

\renewcommand{\shortauthors}{Tony et al.}

\begin{abstract} Conversational agents or chatbots are widely investigated and used across different fields including healthcare, education, and marketing. Still, the development of chatbots for assisting secure coding practices is in its infancy. In this paper, we present the results of an empirical study on SKF chatbot, a software-development bot (DevBot) designed to answer queries about software security. To the best of our knowledge, SKF chatbot is one of the very few of its kind, thus a representative instance of conversational DevBots aiding secure software development. In this study, we collect and analyse empirical evidence on the effectiveness of SKF chatbot, while assessing the needs and expectations of its users (i.e., software developers). Furthermore, we explore the factors that may hinder the elaboration of more sophisticated conversational security DevBots and identify features for improving the efficiency of state-of-the-art solutions. All in all, our findings provide valuable insights pointing towards the design of more context-aware and personalized conversational DevBots for security engineering.
\end{abstract}





\maketitle

\section{\textbf{Introduction}}
\label{sec:intro}




Every year hundreds of organizations and companies around the world are negatively affected by severe data breaches and cyber attacks. To a large extent, this is due to a lack of adequate security measures and controls within information systems. Still, secure coding practices in software projects are far from being the norm as developers often lack of proper training, and security is more of an afterthought than an actual priority \cite{rindell2019managing}. In turn, a large number of threats and vulnerabilities emerge as a result of unsavvy decisions and poor security design practices \cite{cite26}. 

Recent advances in Artificial Intelligence (AI) in general and machine learning in particular have fostered the emergence of chatbots assisting people's decisions in different fronts and domains (e.g., healthcare, education, and marketing) \cite{adamopoulou2020overview}. Certainly, the software industry has not been the exception to chatbot applications, where conversational agents have also being proposed to support core software engineering tasks (e.g., testing and requirements analysis) \cite{cite36}. Moreover, such `DevBots' are slowly becoming popular as well in the field of cyber-security, helping developers to spot security flaws in their~code. 

Several DevBots have been proposed within the current literature to help in the identification of security flaws during development \cite{cite28,cite35,cite36}. These solutions often employ static or dynamic code analysis features for spotting and warning developers about security vulnerabilities. Still, most of these tools do not yet provide \textit{conversational support} to help fixing nor mitigating the bugs they identify \cite{cite30}. Hence, there is a call for more supportive DevBot solutions that could provide instructional content on how to improve the security of software projects through adequate coding practices.

In this work, we aim to delve into the extent to which current conversational DevBot applications assist the development of secure software systems. For this, we conducted an empirical study on SKF chatbot (SKF stands for Security Knowledge Framework), a conversational agent designed to address developers' security-related questions. Particularly, we aimed at evaluating the extent to which the assistance of SKF chatbot is deemed adequate and useful by its users. To gain insights on these performance aspects we ran a study with 15 participants in which they had to use SKF for removing code vulnerabilities. Overall, the research questions (RQs) addressed in this work are the following:

\textbf{\textit{RQ1}: How effective are conversational DevBots for secure programming?} To answer this RQ we analysed the extent to which the DevBot helped participants to remove security vulnerabilities. That is, whether it led to full, partial, or no vulnerability fixes, and compared it against the outcome when no DevBot support was provided.



\textbf{\textit{RQ2}: How useful are conversational DevBots for secure programming?} In this case we developed a questionnaire for capturing the extent to which participants understood the nature of the vulnerabilities they had to fix, and whether they managed to find the right fix for such vulnerabilities. We also assessed whether the DevBot managed to answer all of their questions along with the perceived usefulness of the tool.

\textbf{\textit{RQ3}: Do developers prefer to use a conversational DevBot when conducting security-related tasks?} To answer this RQ we included some questionnaire items assessing developers' intentions towards a DevBot-aided approach over other problem-solving strategies (e.g., a manual online search).

\textbf{\textit{RQ4}: What features do developers expect from a conversational DevBot for secure programming?} This RQ aims to identify features that could improve the effectiveness of current conversational security DevBots. For this we included some open-ended questions by the end of our study to elicit the functionalities of an ``ideal'' chatbot solution.



Our findings suggest that there is still room for improvement in security DevBots, specially when it comes to creating an extensive knowledge base containing reliable information about (i) security vulnerabilities, (ii) fixes, (iii) correct security API usages, and (iv) correct code integration rules. Furthermore, there is a need for more context-aware solutions capable to (i) provide feedback aligned with the individual characteristics of software projects, and (ii) give personalized assistance and support to the end users.


The remaining of this paper is organizes as follows: Section~\ref{sec:rw} gives a brief overview of the existing conversational and security DevBots as well as some background on SKF chatbot, Section~\ref{sec:methodology}
describes our research methodology, the results and their discussion is done in Sections~\ref{sec:results} and \ref{sec:discussion} respectively. We finally conclude the paper and discuss future research directions in Section~\ref{sec:conclusion}.


\section{\textbf{Background and Related Work}}
\label{sec:rw}

In this section we give an overview of the current state-of-the-art in  DevBots development. We also introduce the theoretical foundations for our study, namely SKF chatbot and its main characteristics.

\subsection{Conversational and Security DevBots}

In the recent years, bots and their applications have caught the attention of researchers across different software-related disciplines \cite{cite36}. Specially, there is an increasing interest towards conversational solutions capable of interacting via natural language commands. MSRBot \cite{cite2} for instance is a conversational DevBot supporting tasks related to the mining of software repositories that can address project-specific questions like \textit{``who modified <file\_name>?''} or \textit{``what commits were submitted on <date>?''}. APIBot \cite{cite18} is another conversational agent capable of answering questions about the adequate usage of APIs, thus releasing developers from the burden of scrutinizing multiple pages of API documentation. DevBots incorporating voice-recognition features can also be found within the current literature. Such is the case of Devy \cite{cite16}, a voice-activated assistant helping software practitioners to perform high-level workflow tasks (e.g., submitting code changes for review). Still, research on voice-controlled DevBots is at a very early stage when compared to chat-based support \cite{cite36}.

Security engineering has not been the exception to DevBots and its applications. Particularly, bug identification, automated testing, and code repair are some of the tasks supported by current security DevBots \cite{cite36}. For instance, \citet{wyrich2019towards} proposed a refactoring DevBot capable of spotting code smells and fixing simple warnings. Besides, the bot can also send refactored changes to the developers in the form of pull requests. SAW-BOT \cite{serban2021saw} is another bot helping developers addressing warnings in their code. Like \cite{wyrich2019towards}, SAW-BOT generates fixes that are later suggested to developers via pull requests. Other security DevBots like Repairnator \cite{cite3} have been shaped to support code repair tasks and to monitor test failures in continuous integration. Bots detecting risky commits in open-source projects are also matter of ongoing research efforts~ \cite{cite36}.

\subsection{SKF Chatbot}


Unlike in other software engineering domains, conversational Dev-Bots seem to be underrepresented in the security field. Overall, very few chat-based solutions can be spotted within the current literature,  SKF chatbot\footnote{https://github.com/Priya997/SKF-Chatbot} being one among the most salient ones. SKF was developed as part of OWASP's Secure Knowledge Framework project\footnote{https://www.securityknowledgeframework.org/} to enable quick access to information on security vulnerabilities. It contains a knowledge base of common security weaknesses and vulnerabilities along with a set of code examples. Basically SKF can react to 3 type of inquiries on a particular vulnerability, namely (i) \textit{description}, (ii) \textit{solution}, and (iii) \textit{code snippet}. While the first two provide detailed information in natural language about the vulnerability and its prospective solution, the third one corresponds to a code example either in Django, Java, PHP, Flask or Ruby.

\begin{figure}[hbt!]
    \centering
    \includegraphics[width=\linewidth]{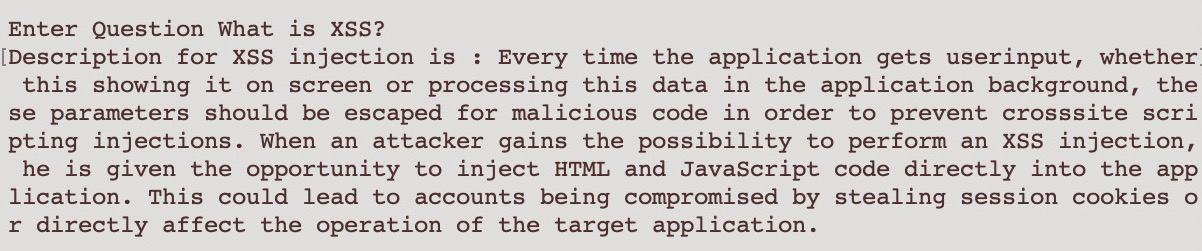}
    \caption{Terminal version of SKF chatbot}
    \label{fig:chatbot}
\end{figure}

SKF chatbot executes three main steps when answering users' queries: (i) intent classification, (ii) entity recognition, and (iii) response generation. In the first step, SKF seeks to understand whether the goal of the user is to get a description, a solution, or a code snippet. For this, it uses a Multinomial Naive Bayes (MNB) classifier to predict the question's intent. Next, it tries to identify the vulnerability for which the user is requesting information. This is done through a keyword-extraction algorithm named Rapid Automatic Keyword Extraction (RAKE), which is domain-independent and helps separating keywords from other words in the question. Finally, SKF generates a response by querying its knowledge base with the intent and the entities extracted from the user's question. \autoref{fig:chatbot} illustrates the output generated for \textit{``What is XSS?''}.

\section{\textbf{Research Methodology}}
\label{sec:methodology}

Despite being one of the few (for not saying the only) security DevBots with conversational features, SKF's usability has not been yet thoroughly investigated to the best of our knowledge. Hence, we conducted and empirical study to analyse its performance, efficiency, and overall user acceptance. In the following subsections we describe the proposed experimental setting along with the instruments used for SKF's assessment.


\subsection{\textbf{Experimental Setup}}

\begin{figure}[hbt!]
    \centering
    \includegraphics[width=\linewidth]{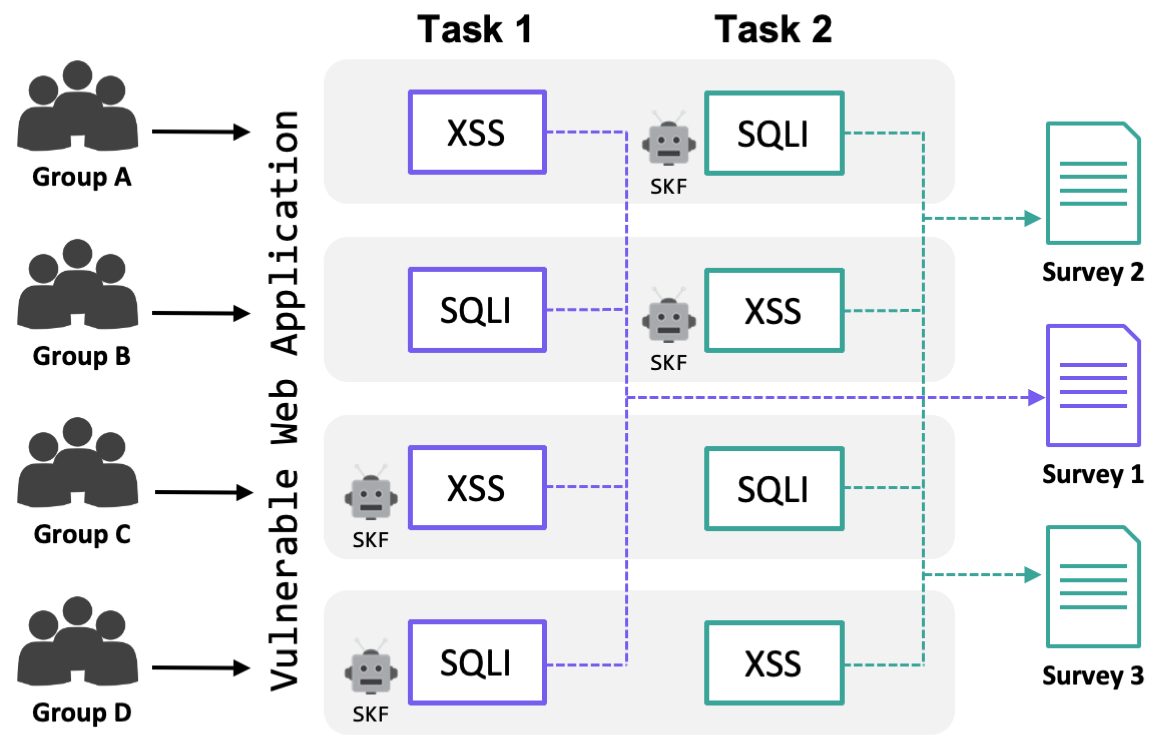}
    \caption{Workflow of the empirical study}
    \label{fig:workflow}
\end{figure}

We conducted the evaluation of SKF chatbot using a Java web application containing 2 code-level security vulnerabilities allowing Cross-Site Scripting (XSS) and SQL Injection (SQLI) attacks. While the former consists of placing of malicious SQL code inside webpage inputs, the latter occurs when malicious code is sent through a web application (usually as browser-side scripts). The participants were given the source code of this web application in which the locations of the vulnerabilities were marked in comments. They were asked to fix the specified vulnerable area of the code with and without SKF's help. For this, we created 4 groups, each receiving the tasks and the DevBot support in a different order to reduce the presence of biases in our analysis (\autoref{fig:workflow}). For instance, participants in \textit{Group A} had to fix the XSS vulnerability first and then the one corresponding to SQLI. In this case, the support of SKF chatbot was given for the resolution of the second task (i.e., SQLI), whereas for the first one (i.e., XSS) they were only allowed to use Internet search.

\subsection{\textbf{Survey Instruments}}

We created 3 survey forms for the experiment containing yes/no, multiple choice, and open-ended questions. Survey  1 was given to the participants after completing the first assignment, whereas Survey 2 and 3 after the second task. Particularly, Surveys 1 and 2 were tailored according to (i) the task the participant had to solve, and (ii) whether she received support of SKF chatbot for its resolution or not. For instance, participants in Group A were asked about the use of online sources for fixing the XSS vulnerability in Survey 1, while in Survey 2 they provided feedback on the use of the conversational DevBot (e.g., their perceived usefulness). Survey 3 remained the same for every group and was given to the participants at the very end of the experiment. Particularly, it contained questions eliciting participants' overall acceptance of the DevBot when solving security vulnerabilities (e.g., if they preferred the DevBot's assistance over a self-conducted Internet search). This third form also included open-ended questions aiming to collect user input that could help shaping additional features for the DevBot. All survey forms and task descriptions were compiled into a \textbf{supplementary~file} available inside a public repository\footnote{Repository link: \url{https://collaborating.tuhh.de/e-22/public/skf-chatbot-empirical-study}.}.

\section{Results} \label{sec:results}

We recruited 15 participants for our experiment (Groups A, B, and C of 4 participants each, and Group D with 3), most of them master students at \textit{*Anonymized Institution*}. In a preliminary assessment, participants' reported having low to medium knowledge on software security, while their general programming skills averaged in medium to high range which made them suitable candidates for evaluating the performance of SKF chatbot. In the following subsections we report our findings in terms of \textit{Effectiveness}, \textit{Q\&A Coverage}, and \textit{Perceived Usefulness}.

\subsection{Effectiveness}

To determine the DevBot's effectiveness, we analysed the number of vulnerabilities successfully fixed by the participants. We considered a vulnerability as ``fixed'' when the correct patch is applied along with the necessary libraries for executing it. When such libraries are not included, the vulnerability is considered as ``partially fixed''. A vulnerability is ``not fixed'' either when no patch is applied or when the wrong solution is implemented. 

\begin{figure*}[hbt!]
    \centering
    \includegraphics[width=0.40\linewidth]{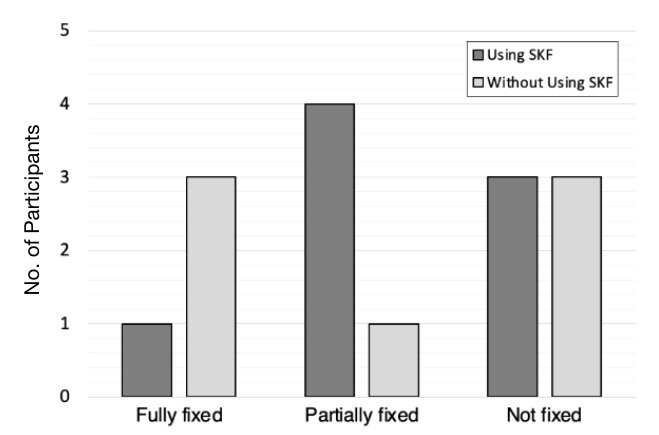}\hspace{2ex}
    \includegraphics[width=0.40\linewidth]{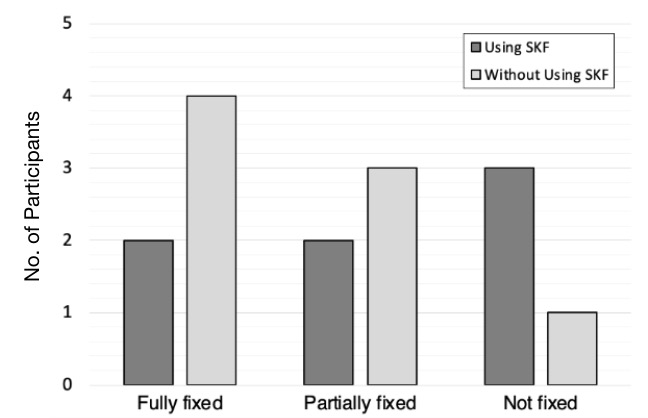}
    \caption{Participant's performance when fixing the XSS (left) and the SQLI vulnerabilities (right).}
    \label{fig:effectiveness}
\end{figure*}

Based on this criterion, we observed that for both vulnerabilities (i.e., XSS and SQLI) the number of full fixes was higher when no DevBot support was provided (\autoref{fig:effectiveness}). Particularly, 7 out of 15 participants managed to fully solve the given task by conducting an Internet search, but only 3 arrived to the right fix when receiving DevBot support. 

When it comes to partial fixes, our results are mixed. As shown in \autoref{fig:effectiveness}-left, there were more participants arriving to a partial solution of the XSS vulnerability with the help of SKF chatbot (4) than using Internet search (1). Conversely, more participants arrived to a partial fix of the SQLI vulnerability without any DevBot support. Particularly, 2 participants arrived to a partial fix with the help of SKF chatbot, whereas 3 did it only by the means of Internet searches (\autoref{fig:effectiveness}-right). The number of unfixed vulnerabilities was for both tasks lower or equal when participants applied a self-conducted Internet search than a DevBot-assisted one.

\subsection{Q\&A Coverage} \label{sec:qacoverage}

Another aspect relevant to the DevBot's performance is the amount of questions it manages to answer. Hence, we asked participants whether SKF chatbot responded to all of their questions to understand to which extent it helped in the resolution of the tasks. All in all, 10 out of 15 participants mentioned that SKF chatbot answered all of their questions, while 3 said it only answered them partially. The remaining 2 participants reported not receiving any useful information from the DevBot for addressing the security vulnerabilities.

To further assess SKF's Q\&A coverage, we asked participants whether they used other resources (e.g., Internet search) in addition to the DevBot for fixing the vulnerabilities. From 15 participants, 9 reported having used additional resources to complete the given task, whereas 6 mentioned that they did not. Among the 9 participants who used extra resources, 5 of them had reported that the DevBot did answer all of their questions.

\subsection{Perceived Usefulness}\label{sec:usefulness}

There are two main steps when removing a vulnerability from a software application, namely (i) understanding the nature of the vulnerability, and (ii) finding the correct fix for such a vulnerability. In order to analyse how well SKF chatbot supports this process, we asked participants whether they would prefer a DevBot over an Internet search on each of these stages or not. 

\begin{figure}[hbt!]
    \centering    \includegraphics[width=0.95\linewidth]{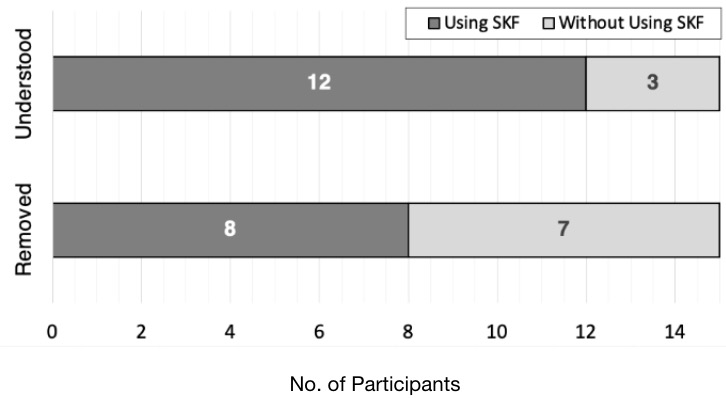}
    \caption{Participants' preferences regarding understanding and removing security vulnerabilities.}
    \label{fig:preferences-graph}
\end{figure}

As shown in Figure \ref{fig:preferences-graph}, 12 out of 15 participants reported that using SKF chatbot helped them to better understand the vulnerability than simply searching for it on the Internet. On the other hand, opinions were more polarized when it comes to finding the right fix. Particularly, 7 out of 15 respondents said they rather look for the correct fix themselves using the Internet, while 8 opted for a DevBot-supported search. Interestingly, only 3 out of 15 participants managed to find full fixes (i.e., either XSS or SQLI) with the help of SKF chatbot. However, if we consider the the total number of full and partial fixes with DevBot support we obtain 9 out of 15, which is closer to the fixing preferences reported by the participants.


Last but not least, we asked a subgroup of participants (8) if they would use a conversational DevBot for addressing security vulnerabilities in the future. The available options were \textit{yes}, \textit{no} and \textit{maybe}. Overall, there was only one participant who was willing to receive DevBot support for applying future security fixes, whereas 5 said they might and 2 that they wont.





\section{\textbf{Discussion}} \label{sec:discussion}

In this section we discuss the results presented in Section~\ref{sec:results} to elaborate on the RQs of our study. This includes aspects related to the performance of SKF chatbot as well as envisaged features for improving chat-based security DevBots. The limitations of our experimental approach are also addressed by the end of this section.

\subsection{Performance (RQ1, RQ2, and RQ3)}

In terms of \textbf{effectiveness} (\textbf{RQ1}), self-conducted Internet searches seem to outperform the DevBot when it comes to full fixes (i.e., 7/15 against 3/15). Moreover, if we consider full and partial fixes as a whole, the SKF chatbot is still outperformed by the use of the Internet (9/15 against 11/15). 
After further inspecting the participants' answers and having a follow-up interview with some of them, we identified some drawbacks in the current DevBot approach. Particularly, participants mentioned that SKF chatbot only reacts to specific input, and performs poorly when asked follow-up questions. Such was the case of Participant 12 (P12) who said:

\begin{quote}
    \textit{P12: to me the bot was not really useful, because I asked different questions and got always the same answer}
\end{quote}

Moreover, one participant mentioned it did not provide any information on the external libraries necessary to implement the prescribed fixes. This last point is manifested in the larger number of partial fixes compared to full fixes. Despite of this limitation, some participants considered the DevBot as a good starting point when solving security vulnerabilities. For instance, P4 reported: 

\begin{quote}
    \textit{P4: For someone who was a beginner at such tasks, I needed more examples with explanation. But it was a good point to start. From the points given by the chatbot I understood what exactly I need to search. }
\end{quote}


When it comes to \textbf{usefulness} (\textbf{RQ2}), most of the participants agreed that SKF chatbot helped them to understand the nature of the vulnerabilities better than the sources they found on the Internet (Section~\ref{sec:usefulness}). In part, this could be because the DevBot reduces the time developers spend retrieving information as it provides a full description of security vulnerabilities, suitable mitigation actions, and code examples. Moreover, unlike code snippets that can be found on the Internet, the ones provided by the chatbot are seen as trustworthy and hence free of additional security vulnerabilities:

\begin{quote}
    \textit{P9: I was looking into other resources to look for specific Java code examples, i.e. API calls. So for me, it was accompanying material, as the chatbot helped me to nail down the problem itself and give me trustworthy information.}
\end{quote}

As mentioned in Section~\ref{sec:qacoverage}, 10 out of 15 participants said that the DevBot answered all their questions. However, some limitations in the Natural Language Processing (NLP) features of the bot were spotted by some of them:

\begin{quote}
    \textit{P9: My question had to match the given question exactly. I ended up with asking for very broad topics and then selecting from the given list of options to find exactly what I was looking for.}
\end{quote}

This issue may be associated with the fact that SKF chatbot uses a multinomial naive Bayes algorithm to predict the intent behind the queries. Nowadays there are much more sophisticated and versatile NLP frameworks that can be used for this purpose, and hence to improve the conversational capabilities of security DevBots. For instance, BERT \cite{cite20} is a novel language representation model which has shown promising results in question-answering and language interface applications. Another suitable approach is the so called Generative Pre-trained Transformer - 3 (GPT-3) \cite{cite21} which uses deep learning to produce human-like text. Such NLP frameworks combined with a larger knowledge base can help shaping more advanced and efficient conversational security DevBots.



As shown in Section~\ref{sec:usefulness}, 6 out of 8 participants were inclined towards a DevBot-aided approach for addressing future security vulnerabilities. Such \textbf{intentions} (\textbf{RQ3}) can be associated with the clarity with which information is presented by the DevBot along with its reliability, accessibility, and promptness: 

\begin{quote}
\textit{P11: The answers (of the DevBot) matched precisely the given problems. In addition the answers were correct, while answers on the Internet are sometimes incoherent and misleading. In a way the Chatbot was a trusted resource, but for explicit code examples I would nevertheless search the internet. } 
\end{quote}

Nonetheless, the DevBot adoption can be hindered due to conversational limitations (as mentioned earlier) and an unpleasant user interface (P11). Even though participants did not make explicit reference to a lack of adaptability to project-specific scenarios, it can also lessen the DevBot's usability to a large extent.



\subsection{\textbf{Envisaged Features (RQ4)}}

Participants also gave their feedback about which \textbf{additional functionalities} (\textbf{RQ4}) would be beneficial for a conversational DevBot addressing security flaws in software projects. Besides from improvements in the UI and conversational features of the bot, they suggested adding security indicators to the prescribed solutions:

\begin{quote}
\textit{P8: Maybe give recommendations like: this solution helps in 90\% of all cases.}\\[1ex]
\textit{P11: Maybe giving it some code lines and an estimation of how secure it is? Perhaps even suggesting possible vulnerabilities.}
\end{quote}


Sorting and displaying the DevBots' prescriptive solutions according to their security level could improve their navigability, and hence ease their selection. Thereby, developers would become aware of the drawbacks/limitations a particular solution may entail (if any). Supporting examples in other programming languages such as Django and Ruby was another feature pointed by one of the participants (P14). A greater support for integrating solutions into specific code projects was also highlighted as a missing functionality of the DevBot (P12). 



Because of our experimental approach, participants did not need to spot the vulnerabilities themselves inside the provided code snippets. Instead, vulnerabilities were given as input so they could conduct the fixes in a controlled experimental fashion. Nevertheless, supporting the detection and identification of security vulnerabilities in software projects should be a key feature of DevBot solutions. Moreover, according to \citet{cite1}, an ``ideal DevBot'' is defined as:

\begin{quote}
    \textit{... an artificial software developer which
is autonomous, adaptive, and has technical as well as social
competence.}
\end{quote}

Hence, conversational security DevBots should also be capable of mimicking the social and technical skills of software developers to create a smooth interaction with their users. 

In a nutshell, our findings suggest that conversational DevBots for software security should incorporate the following envisaged features to their design:

\begin{enumerate}
    \item Strong integration with the development environment.
    \item Automatic code analysis.
    \item Automatic vulnerability detection.
    \item Enhanced natural language interaction.
    \item Context-aware recommendations.
    \item Automatic code repair.
\end{enumerate}

Many of these features require extensive research efforts and interdisciplinary work. Current advances in machine learning, human-computer interaction, and natural language processing can certainly support this quest. Still, a stronger synergy across these disciplines will be necessary to successfully integrate their individual findings into the design of conversational DevBots for security engineering.


    
    
    
    
    
    


\subsection{Limitations}

Although this study has yielded interesting results, there are certain limitations that should be acknowledged. First of all, the size of the sample is relatively small (15) and hence the results may not be generalizable to a larger population. We could also not thoroughly investigate the existence of potential relations between the variables involved in the study (e.g: knowledge level of the users) and the perceived usefulness due to the small sample. Moreover, the study subjects were mostly students (with programming experience) which could also hinder the validity of our findings. Nevertheless, the use of students as participants remains a valid and effective approach for recreating real software engineering settings in laboratory contexts \cite{falessi2018empirical}. Therefore, we have analysed and interpreted the results of our study taking into consideration the limitations given by the sample composition and size.

Another limitation point is related to the survey instruments employed throughout the study. Particularly, some of the questions included in the survey forms may not fully capture the aspects they intended to. For instance, capturing participants' perceived usefulness may require scales containing several survey items. Hence, the outcome of this study should be seen as a first attempt in assessing the performance of conversational security DevBots, and not as ultimate conclusions. In the future we plan to extend our analysis by the means of validated scales and constructs like Technology Acceptance Model \cite{davis1985}, in order to improve the value of our results.

\section{\textbf{Conclusions and Future Work}}
\label{sec:conclusion}


Supporting developers in their programming practices is of utmost importance for the generation of secure software solutions. Recent advances in AI have contributed to the emergence of DevBots aiding security-related tasks. Still, conversational approaches are in their infancy and thus demand further research and scientific insights. In this work, we have shed some light on the limitations and challenges in the development of conversational DevBots for security engineering. Overall, our findings suggest that these technologies are seen as a reliable source of security knowledge. However, they should incorporate features such as context-aware recommendations and automatic code analysis to increase their levels of user acceptance. Particularly, users would largely benefit from a conversational DevBot capable of providing context-specific answers and code examples that can be easily integrated into existing software projects.


A security DevBot of such characteristics would require an extensive Knowledge Base (KB) comprising data on software security vulnerabilities, mitigation methods, code examples, and library dependencies, among others. Nonetheless, curating such a KB by hand can be a tedious task to carry out. Moreover, technical solutions leveraging manually-curated KBs (e.g., CogniCryptGen \cite{cite22}) cover just a small number of security use cases. Alternatively, it could be feasible to elaborate richer KBs by leveraging data from open-source projects available online. GitHub Copilot \cite{cite37} is a commercial tool that employs this approach. However, a recent study shows that 40\% of the code generated by GitHub Copilot may contain security vulnerabilities \cite{cite37}. In the future, we will investigate alternative strategies for constructing KBs suitable for conversational security DevBots, and thereby overcome some of the limitations and challenges presented in this paper. Recently, \citet{fischer2019stack} released a large dataset consisting of security labelled data-code examples extracted from Stack Overflow\footnote{https://stackoverflow.com/}. In future publications we will assess whether such a dataset can be leveraged for expanding the knowledge of security DevBot solutions.



\begin{acks}
This work was partly funded by the European Union's Horizon 2020 programme under grant agreement No. 952647 (AssureMOSS).
\end{acks}

\bibliographystyle{ACM-Reference-Format}
\bibliography{references}


\begin{thebibliography}{21}


\ifx \showCODEN    \undefined \def \showCODEN     #1{\unskip}     \fi
\ifx \showDOI      \undefined \def \showDOI       #1{#1}\fi
\ifx \showISBNx    \undefined \def \showISBNx     #1{\unskip}     \fi
\ifx \showISBNxiii \undefined \def \showISBNxiii  #1{\unskip}     \fi
\ifx \showISSN     \undefined \def \showISSN      #1{\unskip}     \fi
\ifx \showLCCN     \undefined \def \showLCCN      #1{\unskip}     \fi
\ifx \shownote     \undefined \def \shownote      #1{#1}          \fi
\ifx \showarticletitle \undefined \def \showarticletitle #1{#1}   \fi
\ifx \showURL      \undefined \def \showURL       {\relax}        \fi
\providecommand\bibfield[2]{#2}
\providecommand\bibinfo[2]{#2}
\providecommand\natexlab[1]{#1}
\providecommand\showeprint[2][]{arXiv:#2}

\bibitem[Abdellatif et~al\mbox{.}(2020)]%
        {cite2}
\bibfield{author}{\bibinfo{person}{Ahmad Abdellatif}, \bibinfo{person}{Khaled
  Badran}, {and} \bibinfo{person}{Emad Shihab}.}
  \bibinfo{year}{2020}\natexlab{}.
\newblock \showarticletitle{MSRBot: Using bots to answer questions from
  software repositories}.
\newblock \bibinfo{journal}{\emph{Empir. Softw. Eng.}} \bibinfo{volume}{25},
  \bibinfo{number}{3} (\bibinfo{year}{2020}), \bibinfo{pages}{1834--1863}.
\newblock


\bibitem[Adamopoulou and Moussiades(2020)]%
        {adamopoulou2020overview}
\bibfield{author}{\bibinfo{person}{Eleni Adamopoulou} {and}
  \bibinfo{person}{Lefteris Moussiades}.} \bibinfo{year}{2020}\natexlab{}.
\newblock \showarticletitle{An overview of chatbot technology}. In
  \bibinfo{booktitle}{\emph{IFIP International Conference on Artificial
  Intelligence Applications and Innovations}}. Springer,
  \bibinfo{pages}{373--383}.
\newblock


\bibitem[Bradley et~al\mbox{.}(2018)]%
        {cite16}
\bibfield{author}{\bibinfo{person}{Nick~C. Bradley}, \bibinfo{person}{Thomas
  Fritz}, {and} \bibinfo{person}{Reid Holmes}.}
  \bibinfo{year}{2018}\natexlab{}.
\newblock \showarticletitle{Context-aware conversational developer assistants}.
  In \bibinfo{booktitle}{\emph{International Conference on Software
  Engineering, {ICSE}}}, \bibfield{editor}{\bibinfo{person}{Michel Chaudron},
  \bibinfo{person}{Ivica Crnkovic}, \bibinfo{person}{Marsha Chechik}, {and}
  \bibinfo{person}{Mark Harman}} (Eds.). \bibinfo{publisher}{{ACM}},
  \bibinfo{pages}{993--1003}.
\newblock


\bibitem[Brown et~al\mbox{.}(2020)]%
        {cite21}
\bibfield{author}{\bibinfo{person}{Tom~B. Brown}, \bibinfo{person}{Benjamin
  Mann}, \bibinfo{person}{Nick Ryder}, \bibinfo{person}{Melanie Subbiah},
  \bibinfo{person}{Jared Kaplan}, \bibinfo{person}{Prafulla Dhariwal},
  \bibinfo{person}{Arvind Neelakantan}, \bibinfo{person}{Pranav Shyam},
  \bibinfo{person}{Girish Sastry}, \bibinfo{person}{Amanda Askell},
  \bibinfo{person}{Sandhini Agarwal}, \bibinfo{person}{Ariel Herbert{-}Voss},
  \bibinfo{person}{Gretchen Krueger}, \bibinfo{person}{Tom Henighan},
  \bibinfo{person}{Rewon Child}, \bibinfo{person}{Aditya Ramesh},
  \bibinfo{person}{Daniel~M. Ziegler}, \bibinfo{person}{Jeffrey Wu},
  \bibinfo{person}{Clemens Winter}, \bibinfo{person}{Christopher Hesse},
  \bibinfo{person}{Mark Chen}, \bibinfo{person}{Eric Sigler},
  \bibinfo{person}{Mateusz Litwin}, \bibinfo{person}{Scott Gray},
  \bibinfo{person}{Benjamin Chess}, \bibinfo{person}{Jack Clark},
  \bibinfo{person}{Christopher Berner}, \bibinfo{person}{Sam McCandlish},
  \bibinfo{person}{Alec Radford}, \bibinfo{person}{Ilya Sutskever}, {and}
  \bibinfo{person}{Dario Amodei}.} \bibinfo{year}{2020}\natexlab{}.
\newblock \showarticletitle{Language Models are Few-Shot Learners}.
\newblock \bibinfo{journal}{\emph{CoRR}}  \bibinfo{volume}{abs/2005.14165}
  (\bibinfo{year}{2020}).
\newblock
\urldef\tempurl%
\url{https://arxiv.org/abs/2005.14165}
\showURL{%
\tempurl}


\bibitem[Davis(1985)]%
        {davis1985}
\bibfield{author}{\bibinfo{person}{Fred Davis}.}
  \bibinfo{year}{1985}\natexlab{}.
\newblock \showarticletitle{A Technology Acceptance Model for Empirically
  Testing New End-User Information Systems}.
\newblock  (\bibinfo{date}{01} \bibinfo{year}{1985}).
\newblock


\bibitem[Devlin et~al\mbox{.}(2019)]%
        {cite20}
\bibfield{author}{\bibinfo{person}{Jacob Devlin}, \bibinfo{person}{Ming{-}Wei
  Chang}, \bibinfo{person}{Kenton Lee}, {and} \bibinfo{person}{Kristina
  Toutanova}.} \bibinfo{year}{2019}\natexlab{}.
\newblock \showarticletitle{{BERT:} Pre-training of Deep Bidirectional
  Transformers for Language Understanding}. In \bibinfo{booktitle}{\emph{North
  American Chapter of the Association for Computational Linguistics: Human
  Language Technologies, {NAACL-HLT}}}, \bibfield{editor}{\bibinfo{person}{Jill
  Burstein}, \bibinfo{person}{Christy Doran}, {and} \bibinfo{person}{Thamar
  Solorio}} (Eds.). \bibinfo{publisher}{Association for Computational
  Linguistics}, \bibinfo{pages}{4171--4186}.
\newblock


\bibitem[Erlenhov et~al\mbox{.}(2019)]%
        {cite1}
\bibfield{author}{\bibinfo{person}{Linda Erlenhov},
  \bibinfo{person}{Francisco~Gomes de Oliveira~Neto}, \bibinfo{person}{Riccardo
  Scandariato}, {and} \bibinfo{person}{Philipp Leitner}.}
  \bibinfo{year}{2019}\natexlab{}.
\newblock \showarticletitle{Current and future bots in software development}.
  In \bibinfo{booktitle}{\emph{BotSE@ICSE 2019 May 28, 2019}},
  \bibfield{editor}{\bibinfo{person}{Emad Shihab} {and} \bibinfo{person}{Stefan
  Wagner}} (Eds.). \bibinfo{publisher}{{IEEE} / {ACM}}, \bibinfo{pages}{7--11}.
\newblock


\bibitem[Falessi et~al\mbox{.}(2018)]%
        {falessi2018empirical}
\bibfield{author}{\bibinfo{person}{Davide Falessi}, \bibinfo{person}{Natalia
  Juristo}, \bibinfo{person}{Claes Wohlin}, \bibinfo{person}{Burak Turhan},
  \bibinfo{person}{J{\"u}rgen M{\"u}nch}, \bibinfo{person}{Andreas
  Jedlitschka}, {and} \bibinfo{person}{Markku Oivo}.}
  \bibinfo{year}{2018}\natexlab{}.
\newblock \showarticletitle{Empirical software engineering experts on the use
  of students and professionals in experiments}.
\newblock \bibinfo{journal}{\emph{Empirical Software Engineering}}
  \bibinfo{volume}{23}, \bibinfo{number}{1} (\bibinfo{year}{2018}),
  \bibinfo{pages}{452--489}.
\newblock


\bibitem[Fischer et~al\mbox{.}(2019)]%
        {fischer2019stack}
\bibfield{author}{\bibinfo{person}{Felix Fischer}, \bibinfo{person}{Huang
  Xiao}, \bibinfo{person}{Ching-Yu Kao}, \bibinfo{person}{Yannick
  Stachelscheid}, \bibinfo{person}{Benjamin Johnson}, \bibinfo{person}{Danial
  Razar}, \bibinfo{person}{Paul Fawkesley}, \bibinfo{person}{Nat Buckley},
  \bibinfo{person}{Konstantin B{\"o}ttinger}, \bibinfo{person}{Paul Muntean},
  {et~al\mbox{.}}} \bibinfo{year}{2019}\natexlab{}.
\newblock \showarticletitle{Stack overflow considered helpful! deep learning
  security nudges towards stronger cryptography}. In
  \bibinfo{booktitle}{\emph{28th USENIX Security Symposium (USENIX Security
  19)}}. \bibinfo{pages}{339--356}.
\newblock


\bibitem[Koyuncu et~al\mbox{.}(2020)]%
        {cite35}
\bibfield{author}{\bibinfo{person}{Anil Koyuncu},
  \bibinfo{person}{Tegawend{\'{e}}~F. Bissyand{\'{e}}},
  \bibinfo{person}{Jacques Klein}, {and} \bibinfo{person}{Yves~Le Traon}.}
  \bibinfo{year}{2020}\natexlab{}.
\newblock \showarticletitle{FlexiRepair: Transparent Program Repair with
  Generic Patches}.
\newblock \bibinfo{journal}{\emph{CoRR}}  \bibinfo{volume}{abs/2011.13280}
  (\bibinfo{year}{2020}).
\newblock


\bibitem[Kr{\"{u}}ger et~al\mbox{.}(2020)]%
        {cite22}
\bibfield{author}{\bibinfo{person}{Stefan Kr{\"{u}}ger}, \bibinfo{person}{Karim
  Ali}, {and} \bibinfo{person}{Eric Bodden}.} \bibinfo{year}{2020}\natexlab{}.
\newblock \showarticletitle{CogniCrypt\({}_{\mbox{\emph{GEN}}}\): generating
  code for the secure usage of crypto APIs}. In \bibinfo{booktitle}{\emph{{CGO}
  '20: 18th {ACM/IEEE} International Symposium on Code Generation and
  Optimization}}. \bibinfo{publisher}{{ACM}}, \bibinfo{pages}{185--198}.
\newblock


\bibitem[Monperrus et~al\mbox{.}(2019)]%
        {cite28}
\bibfield{author}{\bibinfo{person}{Martin Monperrus}, \bibinfo{person}{Simon
  Urli}, \bibinfo{person}{Thomas Durieux}, \bibinfo{person}{Martin Martinez},
  \bibinfo{person}{Benoit Baudry}, {and} \bibinfo{person}{Lionel Seinturier}.}
  \bibinfo{year}{2019}\natexlab{}.
\newblock \showarticletitle{Repairnator patches programs automatically}.
\newblock \bibinfo{journal}{\emph{CoRR}}  \bibinfo{volume}{abs/1910.06247}
  (\bibinfo{year}{2019}).
\newblock


\bibitem[Nembhard et~al\mbox{.}(2019)]%
        {cite30}
\bibfield{author}{\bibinfo{person}{Fitzroy Nembhard}, \bibinfo{person}{Marco~M.
  Carvalho}, {and} \bibinfo{person}{Thomas~C. Eskridge}.}
  \bibinfo{year}{2019}\natexlab{}.
\newblock \showarticletitle{Towards the application of recommender systems to
  secure coding}.
\newblock \bibinfo{journal}{\emph{{EURASIP} J. Inf. Secur.}}
  \bibinfo{volume}{2019} (\bibinfo{year}{2019}), \bibinfo{pages}{9}.
\newblock


\bibitem[Othmane et~al\mbox{.}(2017)]%
        {cite26}
\bibfield{author}{\bibinfo{person}{Lotfi~Ben Othmane}, \bibinfo{person}{Golriz
  Chehrazi}, \bibinfo{person}{Eric Bodden}, \bibinfo{person}{Petar Tsalovski},
  {and} \bibinfo{person}{Achim~D. Brucker}.} \bibinfo{year}{2017}\natexlab{}.
\newblock \showarticletitle{Time for Addressing Software Security Issues:
  Prediction Models and Impacting Factors}.
\newblock \bibinfo{journal}{\emph{Data Sci. Eng.}} \bibinfo{volume}{2},
  \bibinfo{number}{2} (\bibinfo{year}{2017}), \bibinfo{pages}{107--124}.
\newblock


\bibitem[Pearce et~al\mbox{.}(2021)]%
        {cite37}
\bibfield{author}{\bibinfo{person}{Hammond Pearce}, \bibinfo{person}{Baleegh
  Ahmad}, \bibinfo{person}{Benjamin Tan}, \bibinfo{person}{Brendan
  Dolan{-}Gavitt}, {and} \bibinfo{person}{Ramesh Karri}.}
  \bibinfo{year}{2021}\natexlab{}.
\newblock \showarticletitle{An Empirical Cybersecurity Evaluation of GitHub
  Copilot's Code Contributions}.
\newblock \bibinfo{journal}{\emph{CoRR}}  \bibinfo{volume}{abs/2108.09293}
  (\bibinfo{year}{2021}).
\newblock


\bibitem[Rindell et~al\mbox{.}(2019)]%
        {rindell2019managing}
\bibfield{author}{\bibinfo{person}{Kalle Rindell}, \bibinfo{person}{Karin
  Bernsmed}, {and} \bibinfo{person}{Martin~Gilje Jaatun}.}
  \bibinfo{year}{2019}\natexlab{}.
\newblock \showarticletitle{Managing security in software: Or: How i learned to
  stop worrying and manage the security technical debt}. In
  \bibinfo{booktitle}{\emph{ARES '19}}. \bibinfo{publisher}{ACM},
  \bibinfo{address}{UK}, \bibinfo{pages}{1--8}.
\newblock


\bibitem[Santhanam et~al\mbox{.}(2022)]%
        {cite36}
\bibfield{author}{\bibinfo{person}{Sivasurya Santhanam},
  \bibinfo{person}{Tobias Hecking}, \bibinfo{person}{Andreas Schreiber}, {and}
  \bibinfo{person}{Stefan Wagner}.} \bibinfo{year}{2022}\natexlab{}.
\newblock \showarticletitle{Bots in software engineering: a systematic mapping
  study}.
\newblock \bibinfo{journal}{\emph{PeerJ Computer Science}}  \bibinfo{volume}{8}
  (\bibinfo{date}{02} \bibinfo{year}{2022}), \bibinfo{pages}{e866}.
\newblock


\bibitem[Serban et~al\mbox{.}(2021)]%
        {serban2021saw}
\bibfield{author}{\bibinfo{person}{Dragos Serban}, \bibinfo{person}{Bart
  Golsteijn}, \bibinfo{person}{Ralph Holdorp}, {and} \bibinfo{person}{Alexander
  Serebrenik}.} \bibinfo{year}{2021}\natexlab{}.
\newblock \showarticletitle{SAW-BOT: Proposing Fixes for Static Analysis
  Warnings with GitHub Suggestions}. In \bibinfo{booktitle}{\emph{2021 IEEE/ACM
  Third International Workshop on Bots in Software Engineering (BotSE)}}. IEEE,
  \bibinfo{pages}{26--30}.
\newblock


\bibitem[Tian et~al\mbox{.}(2017)]%
        {cite18}
\bibfield{author}{\bibinfo{person}{Yuan Tian}, \bibinfo{person}{Ferdian Thung},
  \bibinfo{person}{Abhishek Sharma}, {and} \bibinfo{person}{David Lo}.}
  \bibinfo{year}{2017}\natexlab{}.
\newblock \showarticletitle{APIBot: question answering bot for {API}
  documentation}. In \bibinfo{booktitle}{\emph{{IEEE/ACM} International
  Conference on Automated Software Engineering, {ASE}}},
  \bibfield{editor}{\bibinfo{person}{Grigore Rosu},
  \bibinfo{person}{Massimiliano~Di Penta}, {and} \bibinfo{person}{Tien~N.
  Nguyen}} (Eds.). \bibinfo{publisher}{{IEEE} Computer Society},
  \bibinfo{pages}{153--158}.
\newblock


\bibitem[Urli et~al\mbox{.}(2018)]%
        {cite3}
\bibfield{author}{\bibinfo{person}{Simon Urli}, \bibinfo{person}{Zhongxing Yu},
  \bibinfo{person}{Lionel Seinturier}, {and} \bibinfo{person}{Martin
  Monperrus}.} \bibinfo{year}{2018}\natexlab{}.
\newblock \showarticletitle{How to design a program repair bot?: insights from
  the repairnator project}. In \bibinfo{booktitle}{\emph{International
  Conference on Software Engineering: Software Engineering in Practice, {ICSE}
  {(SEIP)} 2018}}, \bibfield{editor}{\bibinfo{person}{Frances Paulisch} {and}
  \bibinfo{person}{Jan Bosch}} (Eds.). \bibinfo{publisher}{{ACM}},
  \bibinfo{pages}{95--104}.
\newblock


\bibitem[Wyrich and Bogner(2019)]%
        {wyrich2019towards}
\bibfield{author}{\bibinfo{person}{Marvin Wyrich} {and} \bibinfo{person}{Justus
  Bogner}.} \bibinfo{year}{2019}\natexlab{}.
\newblock \showarticletitle{Towards an autonomous bot for automatic source code
  refactoring}. In \bibinfo{booktitle}{\emph{2019 IEEE/ACM 1st International
  Workshop on Bots in Software Engineering (BotSE)}}. IEEE,
  \bibinfo{pages}{24--28}.
\newblock


\end{thebibliography}

\end{document}